\documentclass[conference, a4paper]{IEEEtran}
\IEEEoverridecommandlockouts
\usepackage{microtype}
\usepackage[format=plain,justification=centering]{caption}
\usepackage{cite}
\usepackage{stfloats}
\usepackage{amsmath,amssymb,amsfonts}
\usepackage{xcolor}
\usepackage{orcidlink}
\usepackage{physics}
\usepackage{mleftright}
\usepackage{mathtools}
\usepackage{tikz}
\usepackage[T1]{fontenc}

\hypersetup{
   hidelinks,
   pdftitle={Soft Reverse Reconciliation for Discrete Modulations},
   pdfauthor={M. Origlia, M. Secondini},
   pdfsubject={Information Theory},
   pdfkeywords={{information theory}, {reverse reconciliation}, {QKD}, {LDPC}, {PAM}}
}

\def\BibTeX{{\rm B\kern-.05em{\sc i\kern-.025em b}\kern-.08em
    T\kern-.1667em\lower.7ex\hbox{E}\kern-.125emX}}

\title{Soft Reverse Reconciliation for Discrete Modulations%
    \thanks{This work has been partially supported by PNRR MUR project PE0000023-NQSTI, by the QUASAR project and by the SMARTQKD scholarship funded by Sma-RTy Italia SRL and CNR-IEIIT.}
}

\author{
  \IEEEauthorblockN{Marco Origlia\textsuperscript{1,2,3}\orcidlink{0000-0003-4901-4350}
  , Marco Secondini\textsuperscript{1} \orcidlink{0000-0001-5883-5850}
  }
  \IEEEauthorblockA{\textsuperscript{1}\textit{Telecommunications, Computer Engineering and Photonics (TeCIP) Institute} \\
  \textit{Sant'Anna School of Advanced Studies (SSSA)}, Pisa, Italy}
  \IEEEauthorblockA{\textsuperscript{2}\textit{National Council of Research (CNR-IEIIT)}, Pisa, Italy}
  \IEEEauthorblockA{\textsuperscript{3}\textit{Sma-RTy Italia SRL}, Milano, Italy}
}

\newcommand{\matchparen}[1]{\left(#1\right)}
\newcommand{\Abet}[0]{\mathcal{A}}
\newcommand{\Xhat}[0]{\hat{X}}
\newcommand{\xhat}[0]{\hat{x}}

\newcommand{\PP}[1]{P\{#1\}}

\newcommand{\XXhat}[0]{\mathbf{\Xhat}}
\newcommand{\BB}[0]{\mathbf{B}}
\newcommand{\BBhat}[0]{\mathbf{\hat{B}}}

\begin{document}

\maketitle

\begin{abstract}
The performance of the information reconciliation phase is crucial for quantum key distribution (QKD). Reverse reconciliation (RR) is typically preferred over direct reconciliation (DR) because it yields higher secure key rates. However, a significant challenge in continuous-variable (CV) QKD with discrete modulations (such as QAM) is that Alice lacks soft information about the symbol decisions made by Bob. This limitation restricts error correction to hard-decoding methods, with low reconciliation efficiency.
This work introduces a reverse reconciliation softening (RRS) procedure designed for CV-QKD scenarios employing discrete modulations. This procedure generates a soft metric that Bob can share with Alice over a public channel, enabling her to perform soft-decoding error correction without disclosing any information to a potential eavesdropper.
After detailing the RRS procedure, we investigate how the mutual information between Alice's and Bob's variables changes when the additional metric is shared. We show numerically that RRS improves the mutual information with respect to RR with hard decoding, practically achieving the same mutual information as DR with soft decoding. Finally, we test the proposed RRS for PAM-4 signalling with a rate 1/2 binary LDPC code and bit-wise decoding through numerical simulations, obtaining more than 1dB SNR improvement compared to hard-decoding RR.
  
\end{abstract}


\section{Introduction}

The problem of reconciliation has the aim of obtaining 2 identical sequences out of a pair of random correlated data.
This is particularly useful in all applications where these identical sequences are employed as symmetrical keys
and the pair of random correlated data is obtained from data transmission of various natures.
Secrecy is therefore of paramount importance in such kinds of applications.
A possible scenario of shared randomness generation is quantum key distribution (QKD) \cite{Pirandola:20}.

After the first phase, i.e., random data transmission, the reconciliation procedure joins the game to remove all discrepancies in the correlated data pair.
The usual assumption is that the remote parties are provided with a noiseless and public channel for the completion of this procedure.
The privacy amplification phase follows, to purge the information leaked during the previous phases.

When the reconciliation is initiated by the same party who generated and transmitted the random data, namely Alice,
it is called direct reconciliation (DR): after data transmission, Alice generates the syndrome associated with the data
and publicly discloses it. The receiver, namely Bob, will use the channel output and the syndrome to detect and remove the discrepancies with Alice's data.
On the other hand, the reconciliation may be initiated by Bob,
who will use the channel output to produce a syndrome, and possibly some other auxiliary data, to be disclosed to help Alice remove the discrepancies between her data and Bob's received data.
In this case, it is called reverse reconciliation (RR).

When a potential attacker is taken into account, the direction of the reconciliation matters.
In fact, the data potentially leaked during the transmission phase have larger mutual information with the transmitted data than with the received data.
Consequently, the RR exhibits lower information leakage and potentially a higher key generation rate.

For the phase of data transmission, different signalling techniques may be employed.
For instance, in the realm of QKD, the existing protocols may adopt single photons (discrete-variable QKD)
\cite{bennett_quantum_2014}, or coherent states (continuous-variable QKD) \cite{grosshans_continuous_2002}.
Unlike DV-QKD, CV-QKD does not require single photon sources and detectors and can use the same devices and modulation/detection schemes commonly employed in
classical coherent optical communications \cite{lodewyck2005controlling}.

The choice of the information carrier is not only a matter of implementation,
as it also affects the kind of variables with which Alice and Bob are provided at the beginning of the reconciliation procedure.
In the GG02 protocol \cite{grosshans_continuous_2002}, Alice and Bob share correlated Gaussian random variables. 
The most well-known reconciliation procedures for Gaussian variables are slice reconciliation \cite{vanassche2004sliced}
and multi-dimensional reconciliation \cite{leverrier_multidimensional_2008}.

While communication based on Gaussian variables is theoretically well-founded, it poses several practical challenges \cite{jouguet2012analysis}.
Consequently, CV-QKD protocols utilizing discrete constellations have also been explored \cite{leverrier2009unconditional,hirano2017implementation}.
Among these, a particularly promising approach uses probabilistically-amplitude-shaped quadrature amplitude modulation (PAS-QAM), which allows the protocol to closely approach the secure key rate achievable by GG02 \cite{notarnicola2023probabilistic,roumestan2024shaped}.
When the transmitted symbols are drawn from a discrete constellations, DR can be formulated as a classical error correction problem over an additive white Gaussian noise (AWGN) channel, with slight modifications due to Alice and Bob working within an arbitrary coset of a given code. After Alice publicly reveals the particular coset by disclosing the syndrome of the transmitted sequence, Bob uses this syndrome and the soft information available on his side (the AWGN channel output) to infer the most likely transmitted sequence. For this task, efficient codes and soft-decoding algorithms---such as low-density parity-check (LDPC) codes
and belief propagation 
---are available to closely approach channel capacity.
On the other hand, RR---usually preferred over DR, as discussed above---is made more difficult by the lack of soft information on Alice's side. After Bob makes decisions on the received symbols and discloses the corresponding syndrome, Alice has only hard information (the transmitted sequence) to guess Bob's decisions. This limitation restricts error correction to hard-decoding methods, with low reconciliation efficiency. 
For BPSK and QPSK constellations, a soft RR procedure has been proposed by Leverrier \cite{leverrier2009theoretical}, in which Bob discloses the amplitude of the received samples. In this case, the amplitude gives no information to Eve about Bob's decisions (which depend only on the phase), but enables Alice to perform soft decoding as in the DR scheme (and with same efficiency). Unfortunately, this procedure cannot be extended to general discrete constellations (e.g., QAM), for which disclosing the received amplitude would reveal also relevant information about the key.
As per our knowledge, there is no soft RR procedure for general discrete constellations.

In this work, we propose a novel soft RR procedure designed for arbitrary PAM or QAM constellations. The procedure involves Bob disclosing a soft metric that enables Alice to perform soft decoding, without revealing any information to Eve, except for the syndrome, that would be anyways revealed for the purpose of error correction.
After providing a brief overview of the procedure in Sec. \ref{sec:overview}, we dive deep into the formal aspects of the problem in Sec. \ref{sec:theory},
presenting a solution in Sec. \ref{sec:transformation-functions}. 
In Sec. \ref{sec:simulations-results}, we investigate through numerical simulations the performance of the proposed solution. Finally, we draw some conclusions in Sec. \ref{sec:conclusions} .

\section{Reconciliation Softening scheme overview}\label{sec:overview}

\begin{figure*}[b]
  \centering
  \includegraphics[keepaspectratio, width=.75\textwidth,
  trim={0 0 0 0}, clip]{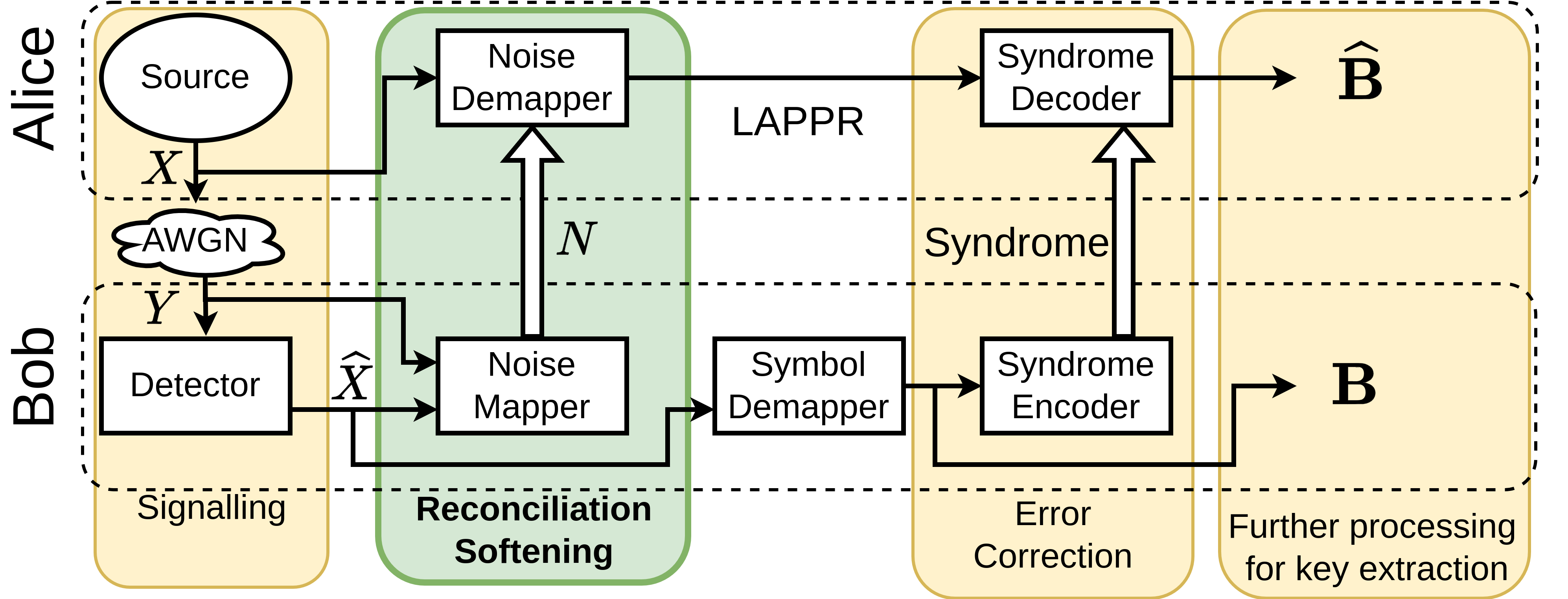}
  \caption{System overview of the scheme. Bob generates a soft metric $N$ from the symbol $Y$ he received \\ and Alice uses it to generate the log-\emph{a posteriori} probability ratios (LAPPR) of the corresponding bits.}\label{fig:system-overview}
\end{figure*}

Fig.~\ref{fig:system-overview} illustrates the working principle of a system employing the proposed reverse reconciliation softening (RRS) procedure. In the initial signalling phase, Alice transmits random symbols $X$ to Bob over an AWGN channel.
Upon receiving  $Y$ as the channel output, Bob makes decisions, denoted as $\Xhat$.
 
Next, the RRS process takes place: Bob uses the channel output $Y$ and the decided symbol $\Xhat$ to generate the side information variable $N$, which he sends to Alice over a noiseless public side channel.
The variable $N$ is intended to provide Alice, who knows $X$, with soft information about Bob's decision $\Xhat$.
At the same time, $N$ is designed to reveal no information about $\Xhat$ to any potential eavesdropper lacking knowledge of $X$, as discussed in Sec.~\ref{sec:information-leakage}.
The functions used to generate this side information are detailed in Sec.~\ref{sec:transformation-functions}.

Alice then uses the side channel information, along with her knowledge of the transmitted symbol, to compute the log-\emph{a posteriori} probability ratios (LAPPRs)
for the bits corresponding to Bob's decisions, as detailed in Sec.~\ref{sec:llr-construction}.

Finally, in the error correction phase, Alice and Bob use an error correction code they have previously agreed upon (for simplicity, we assume here a binary code). After collecting a sufficient number of decisions $\XXhat$, Bob computes the syndrome
of the binary sequence $\mathbf{B}$ corresponding to $\XXhat$ and sends it to Alice over the side channel. Alice feeds the decoder with the syndrome and the LAPPRs to produce an estimate $\mathbf{\BBhat}$ of the binary sequence $\mathbf{B}$.

Further processing follows with the privacy amplification, to remove the information leaked through the syndrome and produce the final shared key.

\section{Theory}\label{sec:theory}
In this section, we focus our analysis on a PAM signalling system. The analysis can be readily applied to QAM signalling consisting of two independent PAMs, one on each quadrature.

Alice, the sender, selects a random sample $X$ from a PAM alphabet $\Abet=\{a_1, \ldots, a_{M}\}$ and sends it over an AWGN channel.
The noise sample added by the channel is $W \sim \mathcal{N}\matchparen{0, \frac{N_0}{2}}$. The noise variance is supposed to be known by Alice and Bob, for instance after channel estimation. The channel output is $Y = X + W$.
Based on $Y$, $N_0$ and $\PP{X=a_i}_{i=1}^{M}$, Bob takes a decision on $\Xhat$
\begin{equation} \xhat = \begin{cases}
    a_1, & y \in D_1 \\
    \vdots & \\
    a_M, & y \in D_M
  \end{cases}
\end{equation}
where the intervals $D_i$ partition the set of real numbers and may be chosen, for instance, to maximize the a posteriori probabilities of the transmitted symbol $X$ (MAP criterion).

Once a sufficient number of decisions have been collected, Bob demaps the resulting sequence $\XXhat$ to a bitstring $\BB$, e.g., assuming a Gray mapping.
Then he evaluates its syndrome
according to a code previously agreed upon, and supposed to be publicly known.
The syndrome is sent to Alice, who will use it to perform the error correction processing and get $\BBhat$, possibly identical to $\BB$.

To assist Alice in the task of decoding $\BBhat$, Bob should share some soft information with her. This information is shared on a public channel.
Since secrecy is the ultimate goal of reverse reconciliation, we devise a procedure to process the available soft information on Bob's side in such a way that it can be shared with Alice without leaking any information on Bob's decisions to a potential eavesdropper.

In our procedure, the soft information shared by Bob with Alice is the realization of a random variable $N$,
which depends deterministically on $Y$.
Without loss of generality, we define the transformation between $Y$ and $N$ through the piecewise function
\begin{equation}\label{eq:functions}
      n = g(y) = \begin{cases}
        g_1(y), & y \in D_1\\
        \vdots & \\
        g_M(y), & y \in D_M.
    \end{cases}
\end{equation}

Given this new random variable $N$, now available at Alice's side along with $X$, we can quantify the knowledge that she has about
Bob's decisions through the mutual information $I(\Xhat;X,N)$, as discussed in Sec. \ref{sec:mutual-information}.
In general, sharing $N$ over a public channel results in an information disclosure about $\Xhat$, which may reduce the secrecy of the final key. The amount of information that can be gained by Eve from the knowledge of $N$ is quantified by the mutual information $I(N;\Xhat)$, assuming that $X$ is not known to anyone but Alice.
Throughout this work we will denote the probability density function of a generic random variable $A$
with $f_A$ and the corresponding cumulative density function with $F_A$.

\subsection{Constraint on information leakage}\label{sec:information-leakage}
In order for $N$ not to reveal any information on $\hat{X}$, we impose the following constraint
\begin{equation}\label{eq:mutual-information-null}
  I(N;\Xhat) = 0 .
\end{equation}
The mutual information between two random variables is zero if and only if they are independent. Therefore, to satisfy (\ref{eq:mutual-information-null}) the function pieces $g_i$ in (\ref{eq:functions}) must be selected such that, given the decision $\Xhat=a_i$, the 
distribution of the resulting variable $N$ is independent of the decision $a_i$ itself, i.e.,
\begin{equation}\label{eq:conditional-distribution-all-equal}
    f_{N|\hat{X}}(n|a_i) = f_{N|\hat{X}}(n|a_j) = f_N(n) \quad \forall\, a_i,\,a_j \in \mathcal{A} .
\end{equation}

\subsection{Transformation functions}\label{sec:transformation-functions}
To find a possible solution to (\ref{eq:conditional-distribution-all-equal}), we further require $g$ to be locally invertible and derivable in each decision interval, i.e., each function piece  $g_i$ is continuously derivable and bijective in $\matchparen{\inf D_i,\;\sup D_i}$.
This requirement will ease the analytical calculation of the LAPPRs (Sec. \ref{sec:llr-construction}).
Furthermore, $g_i$ must yield a random variable with a distribution that does not depend on $i$, so the image of $D_i$ through $g_i$ should not depend on $i$.

We are left with an additional degree of freedom for each function $g_i$, i.e., the sign of the monotonicity---increasing or decreasing. We start by considering the increasing monotonicity, and generalize later for any combination of monotonicity signs of the $g_i$ pieces.

Now we can arbitrarily define a transformation function %
over the decision region of any reference symbol of our choice. %
This way we are fixing the distribution of the output of such function %
and we build all other function pieces in such a way that they satisfy %
the constraint in (\ref{eq:conditional-distribution-all-equal}). %
For our convenience, we choose the transformation function pieces %
such that their outputs are uniformly distributed in $[0,1]$. %
It is immediate to verify that functions of the form %

\begin{equation}\label{eq:gi}
    g_i(y) = \frac{F_Y(y) - F_Y(\inf D_i)}{\Delta F_{Yi}}
\end{equation}
have uniformly distributed outputs in $[0,1]$, {$\Delta F_{Yi} = F_Y(\sup D_i) - F_Y(\inf D_i)$} and, %
for conciseness, {$F_Y(\inf D_i) = \lim_{\tilde{y}\rightarrow (\inf D_i)}F_Y(\tilde{y})$} %
and $F_Y(\sup D_i) = \lim_{\tilde{y}\rightarrow (\sup D_i)}F_Y(\tilde{y})$ (to be used later). %
Consequently functions (\ref{eq:gi}) satisfy constraint (\ref{eq:conditional-distribution-all-equal}).
In other words, the random variables $N|a_i$ for all $i$ are statistically equivalent, 
i. e.,
$\forall i,j: \; N|a_i \cong N|a_j \cong N \sim \mathcal{U}\matchparen{\left[0,1\right]}$.
Each function piece in (\ref{eq:gi}) %
is invertible and derivable in its domain $D_i$
\begin{align}
  &g_i^{-1}(n) = F_Y^{-1}\left(n \cdot \Delta F_{Yi} + F_Y(\inf D_i)\right)\label{eq:invgi}\\
  &g_i'(y) = \frac{f_Y(y)}{\Delta F_{Yi}}
\end{align}
where $f_Y(y) = \frac{1}{\sqrt{2\pi}\sigma}\sum_{a_j \in \Abet}\PP{X=a_j}e^{-\frac{(y-a_j)^2}{2\sigma^2}}$ is the distribution of the channel output $Y$.

It is also immediate to ``flip'' the monotonicity of each single piece of function. In fact, if we want $g_i$ to be monotonically decreasing, we can redefine it as
\begin{align}
  &g_i(y) = \frac{F_Y(\sup D_i) - F_Y(y)}{\Delta F_{Yi}}\\
  &g_i^{-1}(n) = F_Y^{-1}\left(F_Y(\sup D_i) - n \cdot \Delta F_{Yi}\right)\\
  &|g_i'(y)| = -g_i'(y) = \frac{f_Y(y)}{\Delta F_{Yi}}.\label{eq:abs-derivative}
\end{align}

\subsection{Mutual Information}\label{sec:mutual-information}
After describing the construction of the transformation function $g$
in such a way that the constraint (\ref{eq:mutual-information-null}) is satisfied (to avoid information leakage),
we want to quantify the information that Alice, who knows the transmitted symbols $X$,
gains about $\Xhat$ by means of $N$, which is given by $I(\Xhat;X,N)$.

As we will see shortly, the only distribution we explicitly need is $f_{N,\Xhat|X}$. 
We notice that the event \mbox{$\{N=n, \Xhat=a_i\}$}, corresponds to the event \mbox{$\{Y=g_i^{-1}(n)\}$}, and conditionally to the event
\mbox{$\{X=a_j\}$}, its probability distribution is given in (\ref{eq:n-xhat-cond-x}), 
where we substitute the derivative from (\ref{eq:abs-derivative})
\begin{equation}
  \begin{split}
    f_{N,\Xhat|X}(n, a_i|a_j) = \frac{
        f_{Y|X}(g_i^{-1}(n)|a_j)
    }{
        |g_i'(g_i^{-1}(n))|
    }\\
    \label{eq:n-xhat-cond-x}
    = \frac{
        \Delta F_{Yi}
    }{
        \sum_{a_k\in\Abet}\PP{X=a_k}e^{-\frac{(2g_i^{-1}(n) - a_j - a_k)(a_j - a_k)}{2\sigma^2}}
    }.
    \end{split}
\end{equation}
By using (\ref{eq:n-xhat-cond-x}), we can write $f_{N,X|\Xhat}$ as
\begin{equation}\label{eq:n-x-cond-xhat}
    f_{N,X|\Xhat}(n, a_j | a_i)
    = f_{N,\Xhat|X}(n, a_i | a_j) \frac{\PP{X=a_j}}{\PP{\Xhat=a_i}}.
\end{equation}
By marginalizing (\ref{eq:n-xhat-cond-x}) and applying the Bayes theorem, we obtain
\begin{equation}\label{eq:n-x}
  f_{N,X}(n, a_j) = \PP{X=a_j}\sum_{a_i\in\Abet} f_{N,\Xhat|X}(n,a_i|a_j).
\end{equation}
Finally, the mutual information we need is
\begin{equation}
    I(\Xhat;X,N) = \mathop{\mathbb{E}}_{\Xhat,X,N}\left\{\log_2\matchparen{\frac{
        f_{X,N|\Xhat}(X,N|\Xhat)
    }{
        f_{X,N}(X,N)
    }}\right\}
\end{equation}
and by substitution of (\ref{eq:n-x-cond-xhat}) and (\ref{eq:n-x}), it simplifies to
\begin{equation}
    H(\Xhat) + \mathop{\mathbb{E}}_{\Xhat,X,N}\left\{\log_2\matchparen{\frac{
        f_{N,\Xhat|X}(N, \Xhat | X)
    }{
        \sum_{a_k\in\Abet} f_{N,\Xhat|X}(N, a_k | X)
    }}\right\}.
\end{equation}

It can be shown, by applying the chain rule and the data processing inequality, 
that the mutual information of RRS is upper bounded by the direct channel one, i.e.,
\begin{equation}
    I(\Xhat;X,N) \leq I(X;Y).
\end{equation}

\subsection{Construction of the LAPPRs}\label{sec:llr-construction}
Once Alice receives $N$, she can start some guesswork. In particular, since she knows what symbol she transmitted, she can weigh out how likely it is that Bob gives her that particular value $n$ in $M$ different hypothesis.

\begin{figure}[bth]
  \newcommand\AliceX{0}
  \newcommand\BobY{4}
  \newcommand\BobX{5.5}
  \newcommand\SymbBase{-6}
  \newcommand\SymbStep{4}
  \newcommand\AliceSymb{-2}
  \newcommand\LineTop{8}
  \newcommand\LineBottom{-10}
  \newcommand\gaussianScale{5}
  \newcommand\gaussianWidth{0.15}
  \usetikzlibrary{math}
  \resizebox{\columnwidth}{!}{
    \begin{tikzpicture}
      \LARGE
      \draw[gray, thick, dotted] (\AliceX,\LineTop) -- (\AliceX,\LineBottom) node [below, color=black] {$X$};
      \draw[gray, thick] (\BobY,\LineTop) -- (\BobY,\LineBottom) node [below, color=black] {$Y$};
      \draw[gray, thick, dotted] (\BobX,\LineTop) -- (\BobX,\LineBottom) node [below, color=black] {$\hat{X}$};

      \foreach \i in {0,2,3}{
        \tikzmath{
            integer \j;
            \j = \i + 1;
        }
        \fill (\AliceX, \SymbBase+\SymbStep*\i) circle [radius=3pt] node [left]  {$a_\j$};
      };
      \fill [orange] (\AliceX, \AliceSymb) circle [radius=4pt] ;
      \fill (\AliceX, \AliceSymb) node [left] {$x=a_2$};

      \foreach \i in {0,1,2,3}{
        \fill (\BobX, \SymbBase+\SymbStep*\i) circle [radius=3pt];
      };
      
      \foreach \i in {0, 1, 2}{
        \draw (\BobY-0.5,\SymbBase+\SymbStep/2 + \SymbStep*\i) -- (\BobY+0.5,\SymbBase+\SymbStep/2 + \SymbStep*\i);
      }

      \draw[color=orange] plot [variable=\y, domain=\LineBottom:\LineTop, smooth] ({\BobX+2+\gaussianScale*exp(-(\y-\AliceSymb)*(\y-\AliceSymb)*\gaussianWidth)},\y) node [right] {$f_{Y|X}(y|a_2)$};

      \draw[mark options={color=red}] plot [mark=x, mark size=10pt] coordinates {(\BobX,2)} node[right] {$\hat{x}=a_3$};
      
      \foreach \i in {1,2,3}{
        \tikzmath{
            integer \j;
            \j = \i + 1;
        }
        \draw[blue, |->] (\BobY, \SymbBase+\SymbStep*\i) -- (\BobY,\SymbBase+\SymbStep*\i-\SymbStep*0.43) node [left] {$g_\j^{-1}(n)$};
        \draw[blue, ->, dashed] (\AliceX, \AliceSymb) -- (\BobY,\SymbBase+\SymbStep*\i-\SymbStep*0.43);
        \draw[blue] (\BobY,\SymbBase+\SymbStep*\i-\SymbStep*0.43) -- %
        ({\BobX+2+\gaussianScale*exp(-(\SymbBase+\SymbStep*\i-\SymbStep*0.43-\AliceSymb)*(\SymbBase+\SymbStep*\i-\SymbStep*0.43-\AliceSymb)*\gaussianWidth)}, \SymbBase+\SymbStep*\i-\SymbStep*0.43) circle [radius=2pt] ;

      }
      \draw[blue, |->] (\BobY, \SymbBase) -- (\BobY,\SymbBase-0.73*\SymbStep) node [left] {$g_1^{-1}(n)$};
      \draw[blue, ->, dashed] (\AliceX, \AliceSymb) -- (\BobY, \SymbBase-0.73*\SymbStep);
      \draw[blue] (\BobY,\SymbBase-0.73*\SymbStep) -- %
      ({\BobX+2+\gaussianScale*exp(%
        -(\SymbBase-0.73*\SymbStep - \AliceSymb)%
        *(\SymbBase-0.73*\SymbStep - \AliceSymb)*\gaussianWidth)},%
      \SymbBase-0.73*\SymbStep) circle [radius=2pt] ; 

    \end{tikzpicture}
  }
  \caption{Alice hypothesis on $Y$ given the received side information $n$}
  \label{fig:alice-inference}
\end{figure}
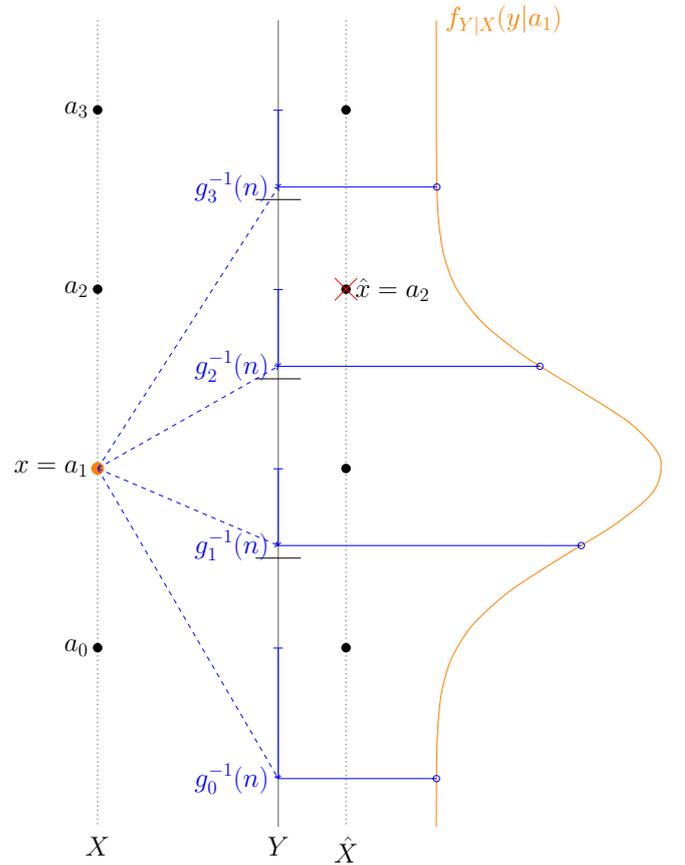

In fact, as sketched in the example of Fig. \ref{fig:alice-inference}, given a particular pair of values $x$ and $n$, the possible events that could have happened are the transitions from symbol $x$ to the $M$ channel outputs $g_i^{-1}(n)$.
As already mentioned at the beginning of section \ref{sec:mutual-information}, the probability density of each of these transition events is given by (\ref{eq:n-xhat-cond-x}), from which we can write the \emph{a posteriori} probability of each possible received symbol as

\begin{equation}\label{eq:app-interval}
    \PP{\Xhat=a_i \; | \; X=a_j, \; N = n} =\\%
    \frac{%
        f_{N,\Xhat|X}(n, a_i| a_j)
    }{%
        f_{N|X}(n| a_j)
    }.
\end{equation}
Subsequently we define the LAPPR on bit $l$ as $\mathcal{L}_l(n, a_j) :=$
\begin{equation}\label{eq:llr}
  \begin{split}
    &\log \matchparen {%
            \frac{
                \sum_{a_i\in\Abet_0(l)} %
                    \PP{\Xhat=a_i \; | \; X=a_j, \; N = n} %
            }{%
                \sum_{a_i\in\Abet_1(l)} %
                    \PP{\Xhat=a_i \; | \; X=a_j, \; N = n} %
            }%
    }\\
    = &\log \matchparen {%
            \frac{
                \sum_{a_i\in\Abet_0(l)} %
                    f_{N,\Xhat|X}(n, a_i| a_j)
            }{%
                \sum_{a_i\in\Abet_1(l)} %
                    f_{N,\Xhat|X}(n, a_i| a_j)
            }%
    }
    \end{split}
\end{equation}
where $\Abet_0(l),\Abet_1(l)\subset\Abet$ partition $\Abet$ according to the value of the $l$-th bit,
given a specific symbol-to-bit demapping rule---Gray (de)mapping, for instance.

\section{Simulations and Results}\label{sec:simulations-results}
\subsection{Mutual information}\label{sec:simulations-results-mi}
In order to assess the performance of the scheme, we compare it against the direct channel and the discrete channel, through their respective mutual information $I(\Xhat;X,N)$, $I(X;Y)$ and $I(X;\Xhat)$.
These curves have been generated numerically, by means of a custom Cython code \cite{Marco_Reverse_Reconciliation_Softening_2024} and the SciPy \cite{2020SciPy-NMeth} Python library.
They have been generated exhaustively for all monotonicity configurations for PAM-4 and BPSK modulation schemes.
Note that for BPSK modulation, i.e., a PAM modulation of order $M=2$, the best configuration is equivalent to Leverrier's scheme\cite{leverrier2009theoretical}, with $|y|$ remapped from $\mathbb{R}^+$ to $[0,1]$, therefore the RRS has the same performance as the direct reconciliation problem (not show in the results).

\begin{figure}[htb]
  \centering
  \includegraphics[keepaspectratio, width=\columnwidth]{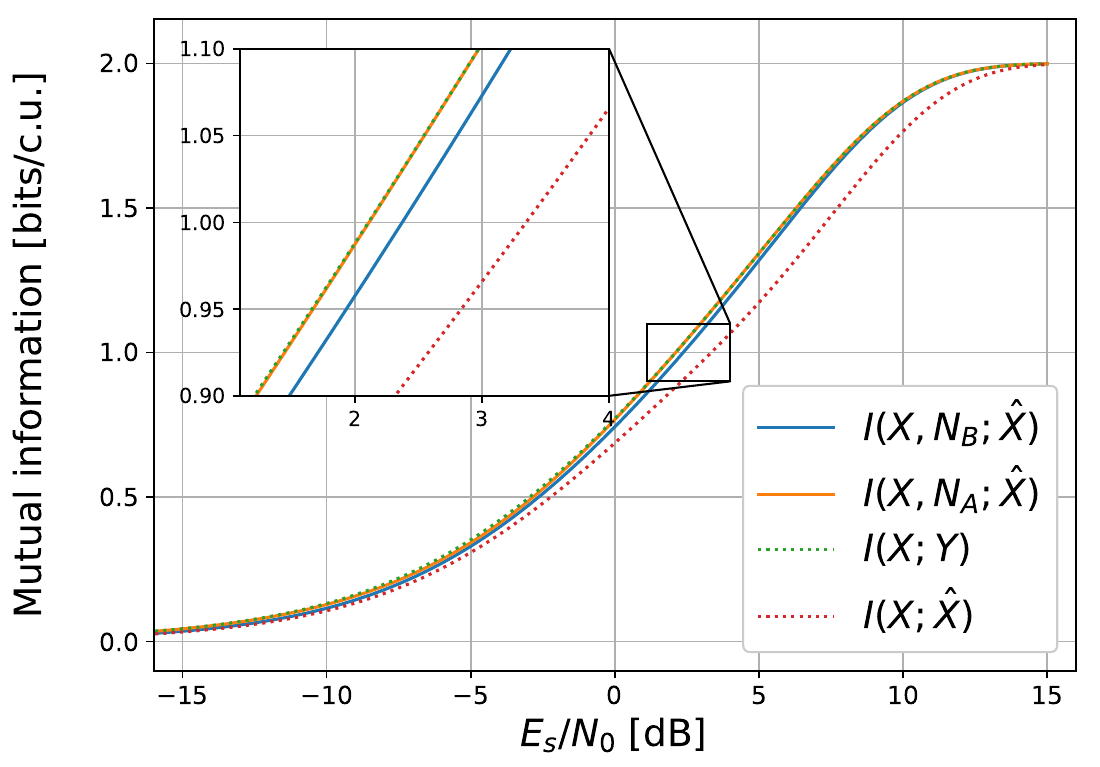}
  \caption{Mutual information for a PAM-4 modulation with different reconciliation directions and configurations.}\label{fig:mutual-information-pam4}
\end{figure}

In Fig. \ref{fig:mutual-information-pam4} we compare two monotonicity configurations against the direct reconciliation and the hard reverse reconciliation in terms of mutual information for a PAM-4 format.
The first monotonicity configuration $C_B$, namely the ``Base'' one, is obtained with all $g_i(y)$ increasing functions.
We will refer to the random variable yielded by these functions with $N_B$,
and to the corresponding mutual information with $I(\Xhat;X,N_B)$.
The best monotonicity configuration $C_A$, namely the ``Alternating'' one, is obtained with alternating monotonicities in adjacent decision intervals.
We will refer to the random variable yielded by this configuration with $N_A$,
and to the corresponding mutual information with $I(\Xhat;X,N_A)$.

\begin{table}[htb]
  \centering
  \begin{tabular}{|l||r|r|r|r|}
    \hline
    $I$ [bit/c.u.] &   Hard &  Base  & Alternating &   Direct\\\hline\hline
    1.75           &   9.85 &   8.65 &        8.56 &   8.56  \\\hline
    1              &   3.35 &   2.37 &        2.11 &   2.11  \\\hline
    0.75           &   0.70 &   0.07 &       -0.17 &  -0.21  \\\hline
    0.3            &  -5.14 &  -5.49 &       -5.71 &  -5.87  \\\hline
    0.1            & -10.26 & -10.61 &      -11.18 & -11.29  \\\hline
    0.01           & -19.99 & -20.29 &      -21.55 & -21.56  \\\hline
  \end{tabular}
  \vspace{1em}
  \caption{SNR $[dB]$ values at which different schemes achieve the given mutual information, for gain comparison.}
  \label{tab:mutual-information-pam4}
\end{table}

To ease the interpretation of the plot of Fig. \ref{fig:mutual-information-pam4},
we report some values from the different curves in Tab. \ref{tab:mutual-information-pam4}.
In this table we rather compare the SNR values at a fixed mutual information value, reported in the first column.
We chose a restricted set of points to better highlight the relative gains in terms of SNR.

For SNR larger than 0 dB, the mutual information of the Alternating configuration mostly overlaps $I(X;Y)$ (for SNR values larger than 1.5 dB, see inset)
; it stays within 0.2 dB for SNR values below 0 dB, again with an overlap for SNR values below -20 dB.
In contrast, the Base configuration has a relative penalty of 0.25 dB at an SNR around 2 dB with respect to the Alternating configuration, although it exhibits a gain larger than 0.6 dB (up to 1.2 dB) for SNR values above 0 dB compared to the hard reverse reconciliation.
Below 0 dB, the mutual information of the Base configuration degrades to the one of the hard reverse reconciliation.

\subsection{Bit Error Rate after Error Correction}\label{sec:simulations-results-ber}

A second set of simulations shows the gain of the application of the described scheme on the coded bit error rate (BER) of a PAM-4 transmission over an AWGN channel. The BER curves have been generated using a DVB-S/2 LDPC code with rate 1/2 and blocklengh 64800. The LDPC decoder employs the belief propagation algorithm and takes into account the syndrome when processing check node inputs to produce the messages for the connected variable nodes. The LDPC decoder and the soft reverse reconciliation functions have been implemented in Cython \cite{Marco_Reverse_Reconciliation_Softening_2024}.

\begin{figure}[hbt]
  \centering
  \includegraphics[keepaspectratio, width=\columnwidth]{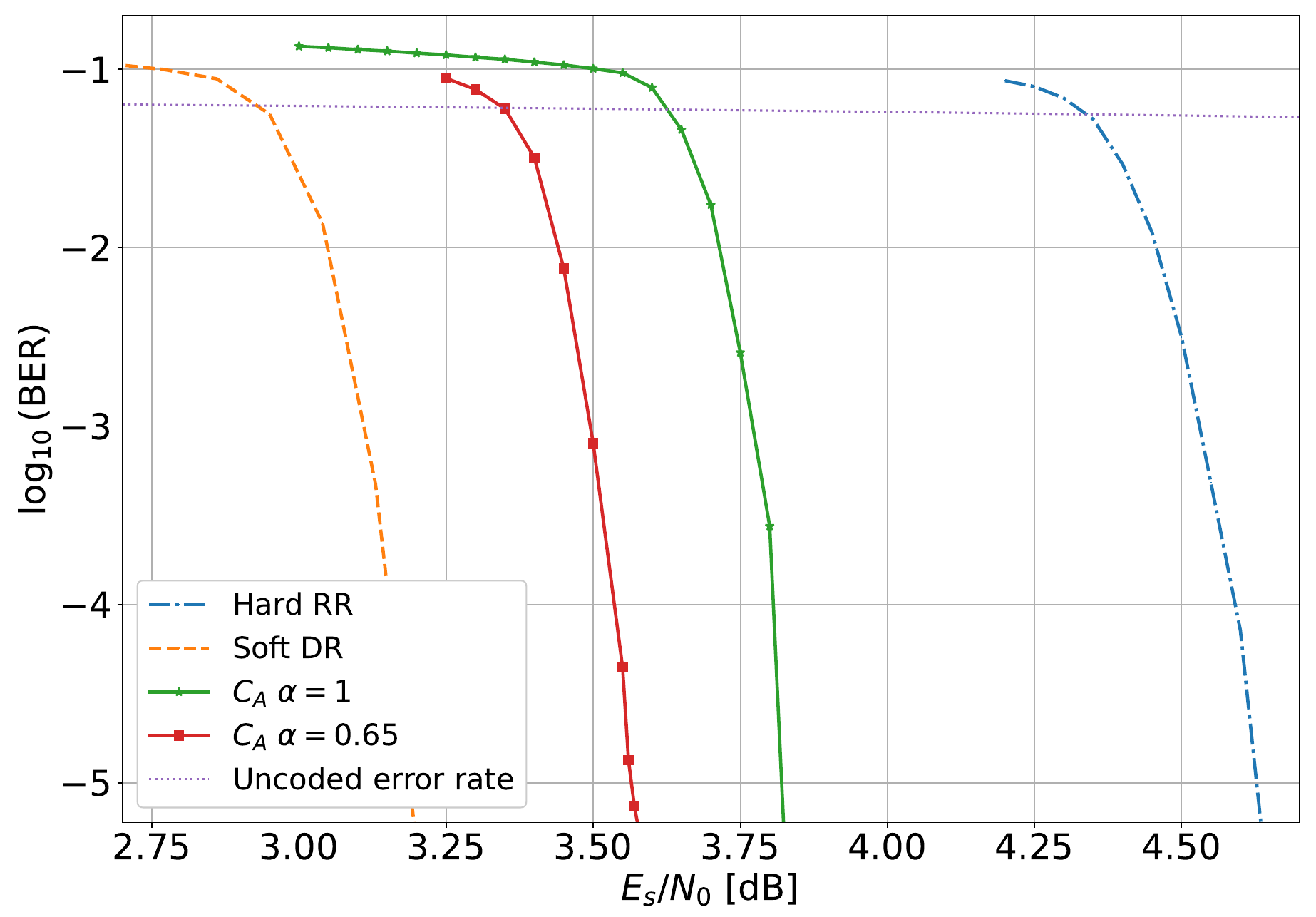}
  \caption{BER characteristic curves for PAM-4, LDPC rate 1/2 and different reconciliation directions and configurations.}\label{fig:ber-pam4}
\end{figure}

In Fig. \ref{fig:ber-pam4} we show the estimated BER characteristics of different reconciliation schemes in the PAM-4 scenario. These curves have been obtained by simulation.

The Alternating configuration $C_A$ exhibits a 0.75 dB gain over the hard reverse scheme.
There is still a gap of 0.65 dB towards the characteristic curve of the direct soft reconciliation,
which is partially covered by multiplying the log-\emph{a posteriori} probability ratios given in Sec. \ref{sec:llr-construction} by a coefficient $\alpha$ 
before feeding them to the decoder. We found that, for a range of values, this coefficient improves the performance of the scheme, 
and the best value obtained by a rough binary search, is around $\alpha=0.65$, 
for which the gain over hard decoding achieves 1\,dB, with a gap of just 0.4\,dB from direct reconciliation.
This gap, which is not present in the mutual information curve, suggests that the selected code is not the optimal choice for this scenario, although it still serves the purpose of highlighting the gain of RRS. The improvement obtained with a non-unitary coefficient $\alpha$ also suggests a possible numerical issue in the calculation of the LAPPRs, which might also be responsible for the performance gap.
The gap, along with the effect of the coefficient $\alpha$, is currently under further investigation.
Nonetheless, in the investigated scenarios, RRS shows an advantage over the hard reverse reconciliation scheme.

\section{Conclusions}\label{sec:conclusions}
In this work, we have introduced the RRS scheme as a method to improve the efficiency of reverse reconciliation for CV-QKD protocols that use discrete modulation alphabets.
This procedure generates a soft metric that helps Alice reconciling her transmitted symbols with Bob's decisions, while not revealing any information about these decisions to a potential eavesdropper.
After formalizing the constraint of zero mutual information between the soft metric and Bob's decisions, we have applied the constraint to design suitable transformation functions $g_i$ for the generation of the soft metric, and derived their inverse for the calculation of the LAPPRs.
Then, we have assessed the performance of the proposed RRS scheme by evaluating the extent to which the mutual information between Alice's symbols and Bob's decisions increases, thanks to the knowledge of the soft metric.
Our scheme is equivalent to Leverrier's reconciliation scheme for the BPSK scenario
\cite{leverrier2009theoretical},
where it achieves exactly the same mutual information as direct reconciliation.
 With a higher modulation order, such as PAM-4, RRS nearly closes the gap between reverse reconciliation based on hard decoding and direct reconciliation based on soft decoding.

This performance is also reflected in the BER curves.
For PAM-4 modulation, RRS achieves a gain of about 1\,dB with respect to the hard reconciliation scheme.
Nonetheless, in contrast to what observed in the mutual information, a small gap of 0.4\,dB still remains with respect to the direct reconciliation curve. The reason for this discrepancy is currently under investigation. It may be due to the non optimal choice of the error correction code, whose selection is not the goal of our investigation, and/or to the use of bit-wise decoding. In fact, as demonstrated in \cite{alvarado_replacing_2015}, mutual information is a good predictor for the post-FEC BER only when non-binary codes are employed, while for bit-wise decoding, BER is better predicted by the normalized generalized mutual information (GMI).

In our future work, we plan to evaluate the GMI of RRS to better understand the impact of the code selection and decoding algorithm on the performance, trying to understand if the gap between hard and soft decoding can be closed also in terms of BER. Furthermore, we will extend the analysis to higher-order modulations, possibly combined with probabilistic constellation shaping, and to different codes and code rates.
Eventually, we will study the performance of the proposed scheme when applied to a practical CV-QKD system.

\bibliographystyle{IEEEtran}
\bibliography{IEEEabrv,references}
\end{document}